\documentclass[aps,prb,twocolumn,showpacs,superscriptaddress,groupedaddress]{revtex4}  % for review and submission
\usepackage{graphicx}  % needed for figures
\usepackage{dcolumn}   % needed for some tables
\usepackage{bm}        % for math
\usepackage{amssymb}   % for math
\usepackage{amsmath}

\begin{document}

% The following information is for internal review, please remove them for submission
\widetext
%\leftline{Version xx as of \today}
%\leftline{To be submitted to PRB}

% the following line is for submission, including submission to the arXiv!!
%\hspace{5.2in} \mbox{Fermilab-Pub-04/xxx-E}

\title{The driving force behind the distortion of one-dimensional monatomic chains -- Peierls theorem revisited}
%\input author_list.tex       % D0 authors (remove the first 3 lines
                             % of this file prior to submission, they
                             % contain a time stamp for the authorlist)
\author{D. Kartoon} 
%\email{danyk@bgu.ac.il}
\thanks{D. Kartoon and U. Argaman contributed equally to this work}
\affiliation{Materials Engineering Department, Ben-Gurion University of the Negev, Beer Sheva 84105, Israel} 
\affiliation{NRCN-Nuclear Research Center Negev, Beer Sheva IL 84190, Israel}

\author{U. Argaman} 
\thanks{D. Kartoon and U. Argaman contributed equally to this work}
\affiliation{Materials Engineering Department, Ben-Gurion University of the Negev, Beer Sheva 84105, Israel}

\author{G. Makov} 
\email{makovg@bgu.ac.il}
\affiliation{Materials Engineering Department, Ben-Gurion University of the Negev, Beer Sheva 84105, Israel}

\date{\today}

\begin{abstract}
The onset of distortion in one-dimensional monatomic chains with partially filled valence bands is considered to be well-established by the Peierls theorem, which associates the distortion with the formation of a band gap and a subsequent gain in energy. Employing modern total energy methods on the test cases of lithium, sodium and carbon chains, we reveal that the distortion is not universal, but conditional upon the balance between distorting and stabilizing forces. Furthermore, in all systems studied, the electrostatic interactions between the electrons and ions act as the main driving force for distortion, rather than the electron band lowering at the Fermi level as is commonly believed. The main stabilizing force which drives the chains toward their symmetric arrangement is derived from the electronic kinetic energy. Both forces are affected by the external conditions, e.g. stress, and consequently the instability of one-dimensional nanowires is conditional upon them. This brings a new perspective to the field of one-dimensional metals, and may shed new light on the distortion of more complex structures.
\end{abstract}

\maketitle

\section{\label{sec:level1}Introduction}
% sections are not used for PRL papers
One-dimensional monatomic chains are of fundamental interest~\cite{Peierls53,Peierls91,McAdon88,Littlewood81}, and a focus of applications in nanotechnology~\cite{Zeng2008,Casillas2014-C,Casari2016-C,Ayuela2002,Cretu2013,LaTorre2015}. 
In his seminal work~\cite{Peierls53,Peierls91}, Peierls argued that one-dimensional evenly-spaced metallic chains can never be stable at zero Kelvin (neglecting the effect of zero-point motion). According to Peierls' theorem, such systems will spontaneously undergo a transition into a more stable lower-symmetry insulating state.   
This transition causes the Fermi surface to coincide with the Brillouin zone boundaries, and it is driven by the opening of an energy gap at the zone boundaries. Peierls showed that the energy gain from such a distortion is
\begin{equation}\label{eq:Peierls}
\Delta E \propto -\tau^2 \cdot log (\tau)
\end{equation}
where $\tau$ is the relative displacement of atoms from their symmetric (equally spaced) positions with a periodicity determined so that the Fermi surface and the edge of the Brillouin zone intersect. At sufficiently small distortions, the energy gain given in Eq.~\ref{eq:Peierls} is always greater than the repulsion between the atomic cores which is assumed to vary as $\tau^2$, thus making the one-dimensional chain inherently unstable. In particular, in the case of half-filled valence bands, the optimal distortion is dimerization of the chain, where every second atom is displaced by $\tau$, and according to Peierls the energy gain is the largest. 

Experiments with one-dimensional carbon chains exhibit dimerization~\cite{Casillas2014-C,LaTorre2015}, often attributed to Peierls distortion. Complex, three-dimensional crystal structures are also often considered to be distorted from higher symmetry lattices due to Peierls-like transition, also refered to as Jones theory~\cite{Jones34,Shick99,Shang2007,Gaspard2016}.  
However, several \textquotedblleft Peierls immune\textquotedblright{} phases have been reported to appear in calculations of one-dimensional chains of different elements, e.g.~\cite{Alemany2009,Sen2006,Khomyakov2006,Ayuela2002,McAdon88}, but attempts to explore their origin are few~\cite{Littlewood81,McAdon88,Johannes2008}. Littlewood and Heine~\cite{Littlewood81} stressed the importance of the electron-electron interactions which were not taken into account in Peierls theorem, arguing that Eq.~\ref{eq:Peierls} is incorrect, but did not depart from its conceptual framework which considers the lowering of energy band at the Fermi level as the main cause of distortion. Johannes and Mazin~\cite{Johannes2008} also analysed a canonical Peierls system of Na atoms and argued that any expected dimerization along the chain axis is energetically unfavourable and doubling of the primitive unit cell occurs only if a transformation into a two-dimensional pattern is allowed. 

In this study we analyse the stability of quasi one-dimensional equally spaced monatomic chains from a total energy point-of-view, in accordance with Peierls. The chains are quasi one-dimensional in the sense that three-dimensional calculations are employed to describe one-dimensional lattices with one-dimensional distortions. We study the associated Born-Oppenheimer energy surface and its component contributions which, to our knowledge, were not studied previously in this context in the framework of density-functional theory (DFT) or similar methods. These modern, more accurate methods may contradict prevailing conceptions and give rise to surprising results.

\section{\label{sec:level1}Methods}
The total energy in Kohn-Sham DFT is given as 
\begin{equation}\label{eq:Ecomps}
E[\rho]=T_{s}[\rho]+E_{xc}[\rho]+E_{Hartree}[\rho]+V_{nn}[\rho]+V_{ne}[\rho]
\end{equation}
where $T_{s}$ is the noninteracting kinetic energy, $E_{xc}$ is the exchange-correlation energy, $E_{Hartree}$ and $V_{nn}$ are respectively the Hartree and Ewald potentials describing the coulombic repulsion between the electrons and between the ions, and $V_{ne}$ is the external potential due to the attraction between the electrons and the ions.
Using the Kohn-Sham formulation, the energy can also be written as~\cite{ParrYang}:
\begin{equation}
\begin{aligned} \label{eq:KSEcomps}
E={} & \sum_{i}^{N}\varepsilon _{i}-\frac{1}{2}\int \frac{\rho (\mathbf{r})\rho ({\mathbf{r}}')}{\left | \mathbf{r}-{\mathbf{r}}' \right |}d\mathbf{r}d{\mathbf{r}}'+E_{Ewald}  \\
     & +E_{xc}[\rho ]-\int v_{xc}(\mathbf{r})\rho(\mathbf{r})d(\mathbf{r}) 
\end{aligned}
\end{equation}
where the first term is the sum over the electronic band energies, the second is minus the Hartree energy to correct for overcounting, the third term is the Ewald energy, and the last two terms are the exchange-correlation adjustment.  

In Peierls' analysis, the electrostatic repulsion is described using an elastic approximation, and the energy gain is due to the lowering of the top of the valence band at the Brillouin zone boundaries evaluated in the nearly-free-electron approximation. Using DFT calculations we can evaluate both the complete electrostatic (classical) contribution (including the attraction between the ions and the electrons) and the overall quantum contribution, which consists mainly of the total kinetic energy of the electrons.  

%\textbf{Methods:}
The electronic structure of the systems considered was calculated using a pseudopotentials plane-waves method and performed with the Quantum Espresso package~\cite{QE}. The exchange-correlation functional was approximated by the PBE general gradient approximation (GGA)~\cite{PBE}. Carbon and sodium pseudopotentials with 4 and 9 valence electrons respectively, were taken from the GBRV database~\cite{GBRV} while lithium pseudopotential with 3 electrons was taken from the PSlibrary database~\cite{PSlibrary}.   
The quasi one-dimensional chains along the z-axis were simulated in tetragonal supercells with vacuum on the x- and y-dimensions, two atoms per unit-cell and periodic boundary conditions. The k-points were aligned homogeneously in the z-direction of the reciprocal lattice vector and centred at the $\Gamma$ point. The calculated systems –- Li, Na and C –- were all set to their equilibrium atomic separation along the z-axis found in our calculations (2.97\AA{}, 3.32\AA{} and 3.83\AA{} respectively) and subjected to one-dimensional distortion which doubled the unit cell along the main axis (dimerization).  

%\textbf{Results:}
\section{\label{sec:level1}Results and discussion}
\begin{figure}
\includegraphics[width=1.0\linewidth]{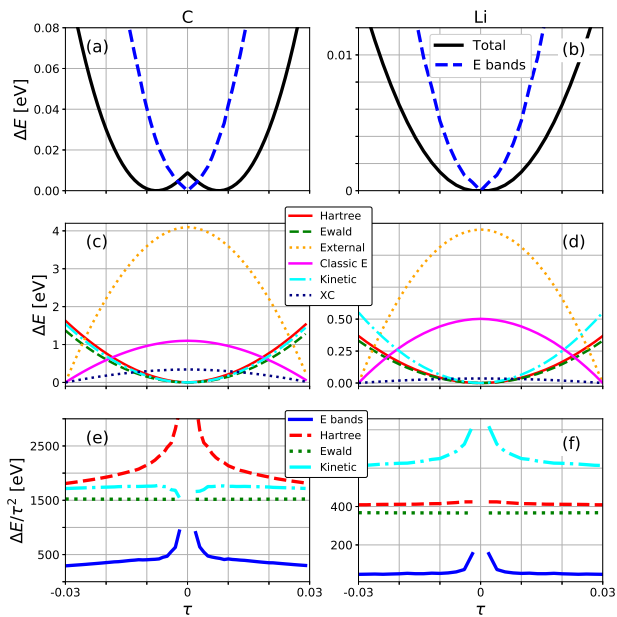}
\caption{\label{fig:CLiTotE} Energy change of carbon and lithium chains due to a one-dimensional distortion. (a)-(b) Total energy and the sum of the electronic band energy (E bands) of carbon (a) and lithium (b) versus distortion parameter $\tau$. (c)-(d) Energy components: Hartree, Ewald, External potential, kinetic and exchange-correlation (XC) of carbon (c) and lithium (d) versus distortion parameter $\tau$. Classic energy denotes the sum of External potential, Ewald and Hartree energies. (e)-(f) Selected energy components divided by $\tau^{2}$ }
\end{figure}

The variation of the energy components with the distortion parameter $\tau$ for lithium and carbon monatomic chains is presented in Fig.~\ref{fig:CLiTotE}. The energy components were taken according to equations \ref{eq:Ecomps} and \ref{eq:KSEcomps} to be the kinetic energy, Hartree, Ewald, external potential and exchange-correlation (XC). In addition Fig.~\ref{fig:CLiTotE} shows the total energy, the sum of the coulomb energies (Classical) and the sum over the electronic band energies (E bands) $\sum_{i}^{N}\varepsilon _{i}$ which is defined as:
\begin{equation}
\sum_{i}^{N}\varepsilon _{i}=\sum_{i}^{N}\left \langle \psi _{i}\left | -\tfrac{1}{2}\nabla^{2}+v_{eff}(\mathbf{r}) \right |\psi _{i} \right \rangle
\end{equation}
where the effective potential is given by:
\begin{equation}
v_{eff}(\mathbf{r})=v_{ext}(\mathbf{r})+\int \frac{\rho ({\mathbf{r}}')}{\left | \mathbf{r}-{\mathbf{r}}' \right |}d\mathbf{r}+v_{xc}(\mathbf{r})
\end{equation}

As can be seen from Fig.~\ref{fig:CLiTotE}a-b, one-dimensional equally-spaced carbon chains are Peierls-unstable as expected and their total energy acquires a double-well shape, while lithium chains are stable, in contradiction to Peierls theorem. However, both systems exhibit the same qualitative behaviour which demonstrates the main problem with Peierls theorem:  in contrast to previous notions, it is apparent that the classical contribution to the overall energy, comprised of the coulombic terms and drawn in Fig.~\ref{fig:CLiTotE}c-d, is the driving force of distortion rather than the quantum energy. In all systems studied, distorted and undistorted, the electrostatic attraction between the electrons and the ions favours the distorted structure and dominates the classical contribution, whereas the kinetic energy which dominates the quantum contribution favours the symmetric lattice.

The overall electronic energy ($E_{bands}$) as plotted in Fig.~\ref{fig:CLiTotE}a-b shows that it is minimized at the equally-spaced chain configuration rather than under distortion. This is one of the most striking results, since the lowering of the bands near the Fermi level is the main cause for distortion according to Peierls, as all other contributions to the energy from changes in the bands far from the Fermi level are neglected. 

The lower panels of Fig.~\ref{fig:CLiTotE} show the behaviour of some of the energy components divided by the square of the distortion parameter versus the distortion parameter $\tau$. According to Peierls' analysis, the electrostatic forces opposing the distortion are represented as an elastic energy which is proportional to $\tau^{2}$, and thus should appear as a horizontal line in these coordinates. However, different contributions to the electrostatic energy, in this case the Ewald and Hartree energies, demonstrate completely different behaviours: The Ewald energy, a sum over an infinite lattice of point-charges, demonstrates a nearly-perfect quadratic dependence on the distortion, whereas the Hartree energy deviates substantially from a $\tau^{2}$ behaviour, resembling qualitatively the $\log{}\tau$ behaviour of the bands energy. This inevitably makes the entire classical energy comprised of the coulombic forces deviate from $\tau^{2}$ behaviour, especially at small distortions, in contrast to Peierls' analysis. A possible explanation for this behaviour is the realistic three-dimensional charge distribution and its distortion which was not originally taken into account. 

\begin{figure}
\includegraphics[width=1.0\linewidth,scale=1]{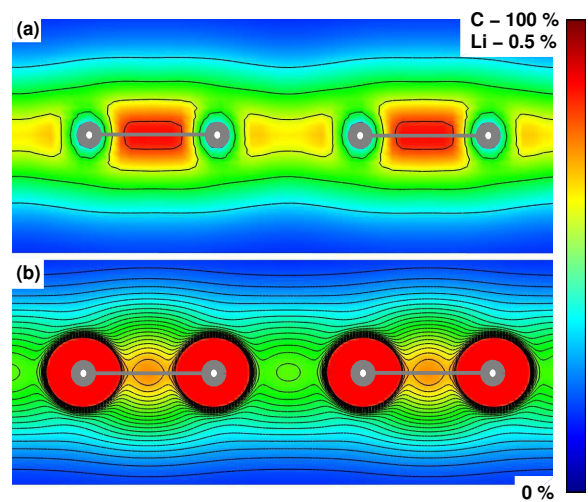}
\caption{\label{fig:ChargeDen} Spatial electronic density distribution for carbon (a) and for lithium (b) linear chains with $\tau=0.03$. Grey spheres represent positions of the ions, and connecting bars are drawn between pairs of nearest neighbours. Colours represent percentage of maximal density value between 0-100\% (carbon) and 0-0.5\% (lithium). The scale for lithium is very narrow (maximal value of 0.5\%, all values above are depicted in red) as most of its charge is centred around the ion cores.}
\end{figure}

To further study the charge distribution we calculate the electronic density, as demonstrated in Fig.~\ref{fig:ChargeDen}. The very different electronic structure of the two chains is apparent: in carbon the highest electron density occurs between neighbouring atoms, whereas in lithium the highest density is distributed in an almost-perfect sphere around each atom. For both carbon and lithium the distribution is not uniform, and the charge density does not spread equally along the chain but rather concentrates between pairs of neighbouring atoms. This charge concentration occurs due to the stronger attraction between the electrons and the ions in the region between adjacent atoms, which increases the energy gain from the interaction between the external potential and the electrons, similar to the function of a bonding orbital in the hydrogen molecule~\cite{Pauling}. 

\begin{figure}
\includegraphics[width=1\linewidth,scale=0.4]{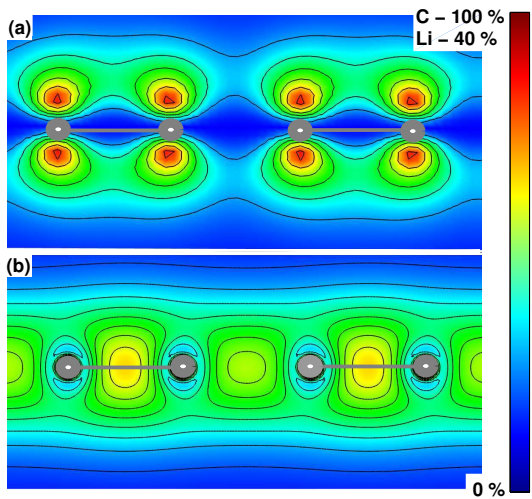}%width=1.0\linewidth,
\caption{\label{fig:ILDOS} Spatial distribution of the integrated local density of states (ILDOS) of the highest occupied band in carbon (a) and lithium (b) linear chains with $\tau=0.03$.  Grey spheres represent positions of the ions, and connecting bars are drawn between pairs of nearest neighbours. Colours represent percentage of maximal ILDOS value between 0-100\% (carbon) and 0-40\% (lithium).}
\end{figure}

The integrated local density of states (ILDOS) of the highest occupied band provides a good description of bonding in real space, especially for lithium where most of the charge density belongs to the lower bands and is concentrated spherically about the ion cores (Fig.~\ref{fig:ILDOS}). In carbon the highest occupied band, corresponding to the $\sigma$-bond of the $p_z$ orbital, is broken into almost discontinuous pairs, whereas in lithium the $\sigma$-bond of the s orbital remains a continuous chain.   

\begin{figure}
\includegraphics[width=1.0\linewidth]{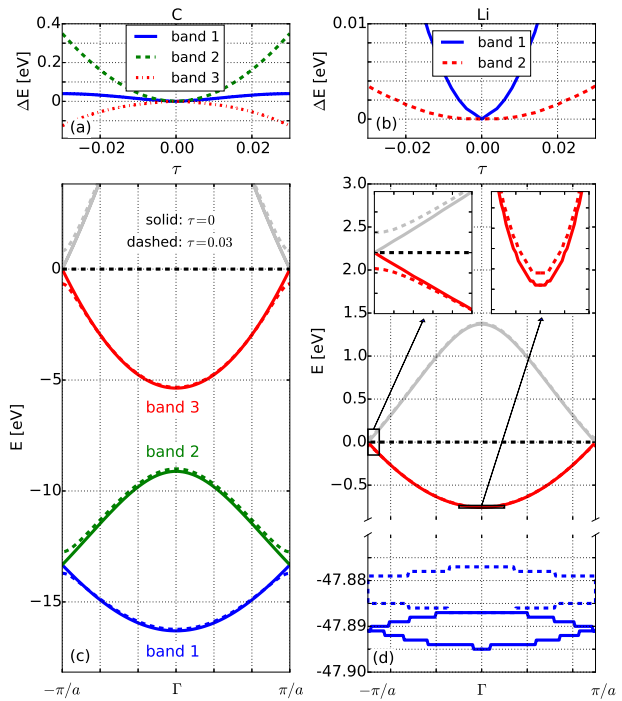}
\caption{\label{fig:Bands} Electronic band energies of carbon and lithium chains. (a)-(b) The energy dependence on distortion of the different bands from the lowest energy band to the highest energy band for carbon (a) and lithium (b). (c)-(d) The band structure of both equilibrium and distorted carbon (c) and lithium (d). Blow-ups of the highest occupied band of lithium at the edge of the Brillouin zone and at the zone center are shown in the insets of (d).}
\end{figure}

In order to understand the behaviour of the band energy with distortion, we analyse in Fig.~\ref{fig:Bands} the electronic band structures of carbon (left) and lithium (right) chains both in their undistorted state and after a 3\% distortion.
As expected, the distortion opens an energy gap at the Fermi level at the edge of the Brillouin zone, although the gap is not symmetric below and above the Fermi level, as demonstrated in the inset of Fig.~\ref{fig:Bands}d. The opening of the gap causes a decrease in the electronic band energy and a metal-to-insulator transition. In carbon, this band-gap suffices to make the highest occupied band energy obtain a maximum value in the undistorted monatomic chain configuration (Fig.~\ref{fig:Bands}a). In contrast, in lithium the changes in the highest occupied band far from the Brillouin zone boundaries (inset of Fig.~\ref{fig:Bands}d) cancel the energy gain and result in a minimum on the equally spaced monatomic chain (Fig.~\ref{fig:Bands}b). In both materials, it is apparent that the distortion significantly affects the bands far below the Fermi level (see Fig.~\ref{fig:Bands}d), resulting in the increase of the overall band energy with distortion (Fig.~\ref{fig:CLiTotE}c and~\ref{fig:CLiTotE}d). 
It was previously observed~\cite{Johannes2008} that in the closely related three-dimensional materials exhibiting a charge density wave, the distortion is not a result of Fermi surface nesting but rather the outcome of combined electronic and ionic interactions in agreement with the present results. These results deviate from the commonly accepted analysis of the band structure in the framework of the nearly-free-electron model used by Peierls, but could be explained within the same framework by using a stronger potential and taking additional higher order terms in the Fourier expansion of the potential.

\begin{figure}
\includegraphics[width=1.0\linewidth,scale=1]{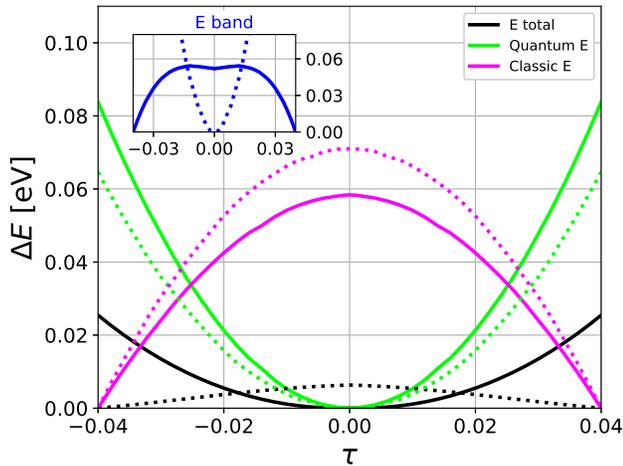}
\caption{\label{fig:Na_strech} Energy change per atom in a sodium one-dimensional chain with distortion $\tau$ for equilibrium lattice parameter a=$3.32\AA$ (solid lines) and stretched lattice parameter a=$3.81\AA$ (dotted lines) corresponding to a tensile stress. Quantum energy is the sum over the kinetic and exchange-correlation energies, classical energy is the sum of Ewald, Hartree and external energies. The sum over the electronic band energies is drawn in the inset.}
\end{figure}

As has been experimentally observed, increasing pressure decreases the distortion (e.g. in carbon~\cite{Cretu2013,Artyukhov2014,LaTorre2015}) and vice versa. Subjecting a stable chain of sodium, for example, to tension, destabilizes it, as was previously reported~\cite{Khomyakov2006} and can be seen in Fig.~\ref{fig:Na_strech}. As the total energy of sodium acquires a double-well shape with stretching, it is insightful to examine its classical (Hartree, Ewald and external) and quantum (kinetic and exchange-correlation) components in order to determine which is the driving force of the instability. It is apparent from Fig.~\ref{fig:Na_strech} that both the classical and the quantum energy components increase their tendency toward distortion with increased lattice parameter. The combination of the classical driving force enhancement with tension and the decrease in resistance to distortion of the quantum energy components, results in the destabilisation of the stretched sodium chain. It is important to note that similarly to carbon, the energy gain does not result from the lowering of the band structure. This is demonstrated clearly in the case of sodium as the total band energy becomes more stable at the symmetric alignment with stretching (see inset in Fig.~\ref{fig:Na_strech}).

\section{\label{sec:level1}Conclusions}
In conclusion, a detailed analysis of all the energy components involved in the dimerization of realistic one-dimensional chains shows that the energy gain is dominated by the coulomb energy, while the energy loss is mainly due to the kinetic energy. The opening of the energy gap on the boundary of the Brillouin zone at the Fermi level is just one contribution to the energy gain which cancels, in some cases, with other contributions from the band structure, including zone-centred states at the valence band and lower energy bands. The instability of one-dimensional chains is found to be dependent on the external stress and not a universal phenomenon as previously considered. This analysis can shed new light on the driving force for more complex distorted structures such as three dimensional structures.

%\input acknowledgement.tex   % input acknowledgement

%\nocite{*}
\bibliographystyle{apsrev4-1} % Tell bibtex which bibliography style to use
\bibliography{Peierls_refs}% Produces the bibliography via BibTeX.

\end{document}